\documentclass{article}

\usepackage{bm}
\usepackage{amssymb}

\usepackage{cite}
\usepackage{amsthm}

\usepackage{bm}
\usepackage{amstext}
\usepackage{psfrag}
\usepackage{amsmath}
\usepackage{amsfonts}
\newcommand{\p}{\partial}
\newcommand{\eu}{{\rm e}}
\newcommand{\dd}{{\rm d}}

\newcommand{\bd}{\begin{definition}}                
\newcommand{\ed}{\end{definition}}                  
\newcommand{\bc}{\begin{corollary}}                 
\newcommand{\ec}{\end{corollary}}                   
\newcommand{\bl}{\begin{lemma}}                     
\newcommand{\el}{\end{lemma}}                       
\newcommand{\bp}{\begin{proposition}}            
\newcommand{\ep}{\end{proposition}}                
\newcommand{\bere}{\begin{remark}}                  
\newcommand{\ere}{\end{remark}}                     

\newcommand{\bt}{\begin{theorem}}
\newcommand{\et}{\end{theorem}}

\newcommand{\be}{\begin{equation}}
\newcommand{\ee}{\end{equation}}

\newcommand{\bit}{\begin{itemize}}
\newcommand{\eit}{\end{itemize}}
\newtheorem{theorem}{Theorem}[section]
\newtheorem{corollary}[theorem]{Corollary}
\newtheorem{lemma}[theorem]{Lemma}
\newtheorem{proposition}[theorem]{Proposition}
\theoremstyle{definition}
\newtheorem{definition}[theorem]{Definition}
\theoremstyle{remark}
\newtheorem{remark}[theorem]{Remark}

\begin{document}

\title{Spacetime metrics from gauge potentials\thanks{Invited contribution to the ``Symmetry'' special issue on ``Physics based on two-by-two matrices.''}}

\author{Ettore Minguzzi \thanks{Dipartimento di Matematica e Informatica ``U. Dini'', Universit\`a
degli Studi di Firenze, Via S. Marta 3,  I-50139 Firenze, Italy E-mail: ettore.minguzzi@unifi.it}}



\date{}

\maketitle

\begin{abstract}
I present an approach to gravity in which the spacetime metric is
constructed from a non-Abelian gauge potential with values in the Lie algebra of the group $U(2)$ (or the Lie algebra of quaternions). If the curvature of this potential vanishes, the metric reduces to a canonical curved background form reminiscent of the Friedmann $S^3$ cosmological metric.
\end{abstract}




\section{Introduction}

The observational evidence in favor of Einstein's general theory of  relativity has clarified that the spacetime manifold is not flat, and hence that it can be approximated by the flat Minkowski spacetime only over limited regions. Quantum Field Theory, and in particular the perturbative approach  through the Feynman's integral, has shown the importance of expanding near a ``classical" background configuration. Although we do not have at our disposal a quantum theory of gravity, it would be natural to take a background configuration  which approximates as much as possible the homogeneous curved background that is expected to take place over cosmological scales accordingly to the cosmological principle. Therefore, it is somewhat surprising that most classical approaches to quantum gravity start from a perturbation of Minkowski's metric in the form $g_{\mu \nu}=\eta_{\mu \nu}+h_{\mu \nu}$. This approach is ill defined in general unless the manifold is asymptotically flat. Indeed, the expansion
 depends on the chosen coordinate system, a fact  which is at odds with the principle of general covariance.

Expanding over the flat metric is like Taylor expanding a function  by taking the first linear approximation near a  point. It is clear that the approximation cannot be good far from the point and that no firm global conclusion can be drawn from similar approaches. A good global expansion should be performed in a different way, taking into account the domain of definition of the function. So, a function defined over an interval would be better approximated with a Fourier series than with a Taylor expansion. Despite of these simple analogies, much research has been devoted to  quantum gravity  by means of expansions of the form $g=\eta+h$, possibly because of the lack of alternatives.

Actually, some years ago \cite{minguzzi02b} I proposed a gauge approach to gravity that solves this problem in a quite simple way and which, I believe, deserves to be better known.

To start with let us observe that general relativity seems to privilege in its very formalism the flat background. Indeed, the Riemann curvature $\mathcal{R}$ measures the extent by which the spacetime is far from flat, namely far from the background
\[
\mathcal{R}=0 \quad \Leftrightarrow  \quad (M,g) \textrm{ is flat}.
\]
If the true background is not the flat Minkowski space then as a first step one would have to construct a different curvature $F$ with the property that
\[
{F}=0 \Leftrightarrow  \ (M,g) \textrm{ takes the canonical  background shape}.
\]
It is indeed possible to accomplish this result. Let us first introduce some notations.


\section{Some notations from gauge theory}

Gauge theories were axiomatized in the fifties by Ehresmann
\cite{kobayashi63} as connections over principal bundles. Since I
need to fix the notation, here I shortly review that setting. A
principal bundle is given by a differentiable manifold (the
bundle) $P$, a differentiable manifold (the base) M, a projection
\begin{equation}
\pi: P \to M
\end{equation}
a Lie group $G$, and a right action of $G$ on P
\begin{equation}
p \to pg \qquad p \in P, \quad g \in G
\end{equation}
such that $M=P/G$, i.e. $M$ is the orbit space.  Moreover, the
fiber bundle $P$ is locally the product $P=M \times G$. To be more
precise, given a point $m \in M$ there is an open set $U$ of $m$,
such that $\pi^{-1}(U)$ is diffeomorphic to $U \times G$ and the
diffeomorphism preserves the right action. If this property holds
also globally the principal bundle is called trivial. The set
$\pi^{-1}(m)$ is the fiber of $m$ and it is diffeomorphic to $G$.
Let $\mathcal{G}$ be the Lie algebra of $G$, and let $\tau_{a}$ be
a base of generators
\begin{equation}
[\tau_{a}, \tau_{b}]=f^{c}_{a b} \tau_{c}.
\end{equation}
Let $p\in P$ be a point of the principal bundle; it can be
considered as an application $p: G \to P$ which acts as $ g \to
pg$. The fundamental fields\footnote{We follow mostly the
conventions of Kobayashi-Nomizu. The upper star $^{*}$  indicates
the pull-back when applied to a function, the fundamental field
when applied to a generator, and the horizontal lift when applied
to a curve or a tangent vector on the base.} $\tau_{a}^{*}$ over
$P$ are defined in $p$ as the push-forward of the group
generators: $\tau_{a}^{*}=p_{*}\tau_{a}$. They are vertical fields
in the sense that they are in the ker of $\pi$:
$\pi_{*}(\tau_{a}^{*})=0$. They form a base of the vertical
tangent space at $p$.

A connection over $P$ is a 1-form $\omega: P \to \mathcal{G}$ with
the following properties
\begin{itemize}
\item[(a)] $\omega(X^{*})=X \qquad X \in \mathcal{G}$
\item[(b)] $R_{g}^{*}\omega=g^{-1} \omega g$
\end{itemize}
The tangent space at $p$ is split into the sum of two subspaces:
the vertical space, that is the ker of $\pi$, and the horizontal
space, that is the ker of $\omega$
\begin{equation}
T_{p}P=H_{p} \oplus V_{p}.
\end{equation}
Let $U$ be an open set of $M$. A section $\sigma$ is a function
$\sigma: U \to \pi^{-1}(U)$ such that $\pi \circ \sigma=I_{U}$.
The gauge potential depends on the section and is defined by
\begin{equation}
A=\tau_{a} A^{a}_{\mu} \dd x^{\mu}=\sigma^{*} \omega
\end{equation}
where $\{ x^{\mu} \}$ are coordinates on the base. A change of
section is sometimes called  gauge transformation. The curvature is defined
by\footnote{The exterior product is defined
through $\alpha \wedge \beta=\alpha \otimes \beta - \beta \otimes
\alpha$ where $\alpha$ and $\beta$ are 1-forms. As a consequence
$\omega \wedge \omega= [\omega, \omega]$}
\begin{equation}
\Omega=d\omega h=\dd \omega + \omega \wedge \omega
\end{equation}
where $h$ projects the vector arguments to the horizontal
space \cite{kobayashi63}. The field strength is defined by
$F=\tau_{a} F^{a}_{\mu \nu} \dd x^{\mu} \dd x^{\nu}=
\sigma^{*}\Omega$. In other words
\begin{equation} \label{mkp}
F^{a}_{\mu \nu}=\p_{\mu} A^{a}_{\nu}-\p_{\nu} A^{a}_{\mu}+f^{a}_{b
c} A^{b}_{\mu} A^{c}_{\nu}.
\end{equation}
Given a section one can construct a system of coordinates over $P$
in a canonical way. Simply let $(x, g)$ be the coordinates of the
point $p=\sigma(x)g$. In this coordinates the connection can be
rewritten
\begin{equation}
\omega=g^{-1}\dd g+g^{-1} A g,
\end{equation}
and the curvature can be rewritten
\begin{equation}
\Omega=g^{-1} F g,
\end{equation}
indeed the form of the connection given here satisfies both the
requirements above and $A=\sigma^{*}\omega$. From these last
equations one easily recovers the gauge transformation rules after
a change of section $\sigma'=\sigma u(x) $ ($ g'=u^{-1}(x) g$),
that is
\begin{eqnarray}
A'_{\mu}&=&u^{-1} A_{\mu} u+u^{-1} \p_{\mu} u \label{gat} \\
F'_{\mu \nu}&=&u^{-1} F_{\mu \nu} u .
\end{eqnarray}
%

\section{The background metric}

We are used to define a manifold through charts $\phi: U\to \mathbb{R}^4$, $U\subset M$, taking values on $\mathbb{R}^4$. Let us instead take them with value in a 4-dimensional canonical manifold with enough structure to admit some natural metric. We shall use a matrix Lie group $G$, but we do not really want to give any special role to the identity of $G$. We shall see later how to solve this problem. The metric $g$ has to be constructed as a small departure from that naturally present in $G$ and which plays the role of background metric.

We take as background metric the expression
\begin{equation}
g_{B} =I_g(\theta,\, \theta),
\end{equation}
where  $\theta$ is the Maurer-Cartan form of the group \cite{kobayashi63}, that is $\theta=g^{-1}\dd g$, and $I_g$ is an adjoint invariant quadratic form on the Lie algebra $\mathcal{G}$, which might depend on $g\in G$. The Maurer-Cartan form has the effect of mapping an element $v\in T_gG$ to the Lie algebra element whose fundamental vector field at $g$ is $v$.

Of course, we demand that $g_B$ be a  Lorentzian metric in a four-dimensional Lie group, and furthermore we want it to represent an isotropic cosmological background, thus $G$ has to contain the $SO(3)$ subgroup.
  We are  lead to the  Abelian group of translations $T_4$ or to the group $U(2)$ (or equivalently the group of quaternions since it shares with $U(2)$ the Lie algebra). In what follows we shall only consider the latter group, the case of the Abelian translation group being simpler.

Thus let us consider the group $U(2)$. Every matrix of this group reads $u=e^{i\lambda} r$ with $0\le \lambda \le \pi$ where $r\in SU(2)$ (while a quaternion reads $e^\lambda r$, $\lambda\in \mathbb{R}$)
\begin{equation}
r=\left( \begin{array}{cc} r_{0}+ir_{3} & r_{2}+ir_{1} \\ -r_{2}+ir_{1} & r_{0}-ir_{3} \end{array} \right) \, , \qquad \sum_{\mu=0}^{3} r_{\mu}^{2}=1.
\end{equation}
The Lie algebra of $U(2)$ is that of anti-hermitian matrices $A$  which read
\begin{equation}
{A}=i\left( \begin{array}{cc} a^{0}+a^{3} & a^{1}-ia^{2} \\ a^{1}+ia^{2} & a^{0}-a^{3} \end{array} \right).
\end{equation}
By adjoint invariance of $I_g$ we mean $ I_{u'g{u'}^\dagger}(u A u^{\dagger},u A u^{\dagger})=I_g(A,A)$, for any $u,u'\in U(2)$. Clearly, the adjoint invariance for the Abelian subgroup $U(1)$ is  guaranteed because for $u\in U(1)$, $u Au^\dagger=A$, $u'g {u'}^\dagger=g$.
 The expressions that satisfy this invariance property are
\begin{equation}  \label{metrica}
I_g(A,A)=\frac{\alpha(\lambda)}{2} ({\rm tr} \, A)^2 -\frac{\beta(\lambda)}{2} {\rm tr} (A^2)
\end{equation}
\begin{equation}
I_g(A,A)=-2 \alpha(\lambda) (a^{0})^{2}+\beta(\lambda)[(a^{0})^2+(a^{1})^{2}+(a^{2})^{2}+(a^{3})^{2}].
\end{equation}
where $\alpha$ and $\beta$ are functions of the phase of $g=e^{i\lambda } r$, $r\in SU(2)$ (which is left invariant under adjoint transformations). We get a Lorentzian metric for $ 2\alpha > \beta$ and $\beta >0$.
With the simple choice $\alpha=\beta=1$ we get
\begin{equation}
I_g(A,A) ={\rm det} A=  -(a^{0})^{2}+(a^{1})^{2}+(a^{2})^{2}+(a^{3})^{2}.
\end{equation}
Notice that  $ {\rm tr} (r^{\dagger} \dd r)=0$ and
\begin{equation}
{\rm tr} (r^{\dagger} \dd r \,r^{\dagger} \dd r)=-{\rm tr} (\dd r^{\dagger} \dd r)=-2 \, {\rm det}(r^{\dagger} \dd r)=-2 \sum_{\mu=0}^{3} \dd r_{\mu}^{2}.
\end{equation}
Let us recall that $\theta=\phi^{\dagger} \dd \phi$ where the group element $\phi$ reads $\phi=r \eu^{i \lambda }$. Thus using $ {\rm tr} (r^{\dagger} \dd r)=0$ we find for the background metric
\begin{align*}
 g_B=I_g(\theta,\theta)  &=  I\left(r^{\dagger} \dd r+ i\dd \lambda, r^{\dagger} \dd r+ i\dd \lambda\right)= \nonumber\\
& = -I\left(\dd \lambda,\dd \lambda  \right)+ I(r^{\dagger} \dd r,r^{\dagger} \dd r)
 = -(2 \alpha -\beta) \dd \lambda^{2}- \frac{\beta}{2} {\rm tr} (r^{\dagger} \dd r \,r^{\dagger} \dd r)= \nonumber\\
 & = - (2\alpha-\beta)  \dd \lambda^{2}+\beta (\dd r_{0}^{2}+\dd r_{1}^{2}+\dd r_{2}^{2}+\dd r_{3}^{2})
\end{align*}
Recalling the constraint $\sum_{\mu=0}^{3} r_{\mu}^{2}=1$ we find a background metric which coincides with Friedmann's with a $S^3$ section.

More specifically, let $\sigma_0=I$, and let $\sigma_i$, $i=1,2,3$, be the Pauli matrices. Let $\tau_\mu=i \sigma_\mu$ be a base for the Lie algebra of $U(2)$.
Let us parametrize $\phi\in U(2)$ through
\begin{equation}
\phi=e^{i\lambda \sigma_0} r=e^{i\lambda \sigma_0}  \eu^{i\chi(\tau_{1} \sin \theta \cos \varphi +\tau_{2} \sin \theta \sin \varphi +\tau_{3} \cos \theta)} \ ,
\end{equation}
then the background metric reads
\begin{equation}
 g_B=-\dd t^{2}+a^{2}(t) \left( \dd \chi^{2}+\sin^{2}\chi (\dd \theta^{2}+\sin^{2} \theta \, \dd \varphi^{2}) \right),
\end{equation}
where
\begin{equation} \label{bigbang}
t=\int_{0}^{\lambda} \dd \lambda' \sqrt{2 \alpha(\lambda') -\beta(\lambda')} \ ,
\end{equation}
and
\begin{equation}
a^{2}(t)=\beta \left(\lambda(t) \right).
\end{equation}
These calculations, first presented in \cite{minguzzi02b},   show that the Friedmann metric appears rather naturally  from the study of the $U(2)$ group. Of course, since this argument  depends only on the Lie algebra rather that the group structure, it can be repeated for the group of quaternions \cite{trifonov07}.

\section{Perturbing the background}

In this section we shall suppose that $I_g$ does not depend on $g$, namely that $\alpha$ and $\beta$ are constants, this means that we ignore the time dependence of the cosmological background.

We mentioned that we wish to use charts $\phi: U\to G$, $U\subset M$, with value in a group $G$ but that we do not want to assign to the identity of $G$ any special role.
To that end, let us assume for simplicity that $M$ is simply connected, and let us introduce a trivial bundle $P$ endowed with a flat connection $\tilde{\omega}$. The connection being flat is integrable, thus given an horizontal  section $\tilde{\sigma}\colon M \to P$, and parametrizing every point of $P$ through $p(x,g)=\tilde{\sigma}(x) g$, we obtain a splitting $P\sim M \times G$.
In this way the identity of $G$ does not play any special role since it refers to different points of $P$ depending on the choice of section $\tilde{\sigma}$.

A second section $\sigma\colon M \to P$ is now related to the former by $\sigma(x) \phi^{-1}(x)=\tilde{\sigma}(x)$, where $\phi\colon M \to G$ is the chart we were looking for.  In order to be interpreted as a chart, $\phi$ has to be injective.
The idea is to define the metric
\[
g=I(\tilde{A}-A,\tilde{A}-A) ,
\]
where $\tilde{A}=\sigma^*\tilde{\omega}$ is the potential of the flat connection and $A=\sigma^{*}\omega$ is the potential of a possibly non-trivial connection. From the transformation rule for the potential (\ref{gat}) we obtain
\[
\tilde{A}=\phi^{-1}(x) \, \dd \phi(x).
\]
Let us show that the metric so defined satisfies the property $F=0 \Rightarrow$ background metric.
 Suppose that $F=0$ then $\sigma$ can be chosen in such a way that $A=0$, thus the metric becomes
\begin{equation}
F=0 \quad \Rightarrow \quad  g=I(\phi^{-1}(x) \dd \phi(x),\phi^{-1}(x) \dd \phi(x))=I(\phi^*\theta,\phi^*\theta)=\phi^* g_{B}
\end{equation}
that is, up to a coordinate change the metric coincides with the background metric.

We observe that $A=\tau_a A^a_\mu \dd x^\mu$ has 16 components, namely the same number of components  as the metric. However, we have an additional degree of freedom given by $\phi(x)$. This function can be completely removed using the invertibility of this map, namely using the coordinates $\phi^\mu$ on the Lie group to parametrize  $M$. In this way the metric reads
\[
g=I(\phi^{-1}\dd \phi-\tau_a A^a_\mu(\phi)\dd \phi^\mu,\phi^{-1}\dd \phi-\tau_a A^a_\mu(\phi)\dd \phi^\mu);
\]
these coordinates are referred as {\em internal coordinates}.
In internal coordinates any gauge transformation induces a coordinate transformation. For instance, the gauge potential transforms as
\begin{equation} \label{inttrasf}
\tau_{a} {A'}^{a}_{c} =\{u^{-1}\tau_{a} A^{a}_{b}u+u^{-1} \p_{b}
u\} \frac{\p \phi^{b}}{\p {\phi'}^{c}}\, ,
\end{equation}
and the transformation law for the curvature becomes
\begin{equation}
F'_{a b}=u^{-1}F_{c d}\, u \,\frac{\p \phi^{c}}{\p {\phi'}^{a}}
\frac{\p \phi^{d}}{\p {\phi'}^{b}}\, .
\end{equation}
where $\sigma'=\sigma u$ and the matrix $u(\phi)$ is related to the transformation
${\phi'}^{ a}(\phi^{b})$ by the product $\phi'=\phi \,u(\phi)$. In
the same way it can be shown, for example,  that the spacetime metric  transforms as a tensor under (\ref{inttrasf}).

One can further ask whether  the Einstein equations can be rephrased as
dynamical equations for the potential $A$. The answer is affirmative and passes through the vierbein reformulation of the Einstein-Hilbert Lagrangian.

We recall that a tetrad field (vierbein) $e_a=e^\mu_a \p_\mu$, is a set of four vector fields $e_a$ such that
$g_{\mu \nu}=\eta_{a b}e^{a}_{\mu}e^{b}_{\nu}$. The inverse $e^a_\mu$ is defined through $e^\mu_a e^a_\nu=\delta^\mu_\nu$. The Einstein Lagrangian can be rewritten
\begin{equation} \label{riemmaniana}
-\frac{\sqrt{-g}}{16 \pi} R=\frac{1}{8 \pi}v^{\nu}_{, \,
\nu}+\frac{\sqrt{-g}}{16 \pi} \left\{\frac{1}{4} C^{a b c} C_{a b
c}-C^{a}_{\ a c} C^{b \ c}_{\ b}+\frac{1}{2}C^{a b c}C_{b a c}
\right\} \, ,
\end{equation}
where the first term on the right-hand side is a total divergence  and
\begin{equation} \label{comm.}
C^{c}_{ab}=e^{c}_{\alpha}(\p_{a}e^{\alpha}_{b}-\p_{b}e^{\alpha}_{a})
=e^{\mu}_{a}e^{\nu}_{b}(\p_{\nu}e^{c}_{\mu}-\p_{\mu}e^{c}_{\nu}) \, .
\end{equation}

In order to obtain a dynamics for $A$ we select a base $\tau_a$ for the Lie algebra such that
\[
I(\tau_a,\tau_b)=\eta_{ab},
\]
where $\eta_{ab}$ is the Minkowski metric. Then we make a gauge transformation so as to send the flat potential $\tilde{A}$ to zero. This gauge is called the {\em OT gauge}. Since $g=I(\tau_a A^a_\mu \dd x^\mu,\tau_a A^a_\mu \dd x^\mu)$, the vierbein becomes coincident with the potential
\[
e^a_\mu= A^a_\mu,
\]
so the field equations can be ultimately expressed in terms of $A^a_{\mu}$. We have observed above that with $F=0$ the metric becomes that of the
Einstein static Universe which is not a solution of the dynamical equations (without cosmological constant).
One could wish to obtain a realistic cosmological solution for $F=0$. At the moment I do not known how to modify the theory so as to accomplish this result (but observe observe that we never changed the dynamics which is always that given by the Einstein's equations). However, our framework might not need any modification. It can be shown \cite{minguzzi02b} that the scale factor $a$ in front of the Einstein static Universe metric is actually the coupling constant for this theory so the expansion of the Universe could be an effect related to the renormalization of the theory.

%
%

In the Abelian case $T_4$ (not in the $U(2)$ case) the Lagrangian can also be expressed in terms of the curvature (\ref{mkp}).
Indeed, since $f^c_{ab}=0$ the curvature becomes coincident with the tensors $C^a_{bc}$ entering the above expression of the Lagrangian (however, observe that the potential still enters the metric and the vierbeins which are used to raise the indices of the curvature).
The final expression is  quadratic in the curvature $F$ and is related to the teleparallel approach to general relativity \cite{cho76,hayashi79,rodrigues07,aldrovandi13}.
Issues related to the renormalizability of the dynamics determined by (\ref{riemmaniana}) have yet to be fully studied.

The {\em OT gauge} approach has been used to infer the dynamics and is complementary to the {\em internal coordinates} approach mentioned above. Indeed, while the latter allows us to interpret the map $\phi: U \to G$, $U\subset M$, as a chart with values in $G$, the {\em OT frame} approach sends $\phi$ to the identity, so in the new gauge the non-injective map $\phi$ cannot be interpreted as a chart. Thus, after having developed the dynamics in the {\em OT gauge} we would have to make a last gauge transformation to reformulate it in internal coordinates.


\section*{Acknowledgments}
This work has been
partially supported by GNFM of INDAM.






\end{document}